\documentclass[10pt, preprint]{emulateapj}

\usepackage{color}

\usepackage{style_wei_symbols}
\usepackage{style_wei}

\usepackage{natbib}
\begin{document}


\title{Direct Imaging by {\it SDO} AIA of Quasi-periodic Fast Propagating Waves 
of $\sim$$2000 \kmps$ in the Low Solar Corona}





\author{Wei Liu\altaffilmark{1,}\altaffilmark{2}, Alan M.~Title\altaffilmark{1}, Junwei Zhao\altaffilmark{2}, 
        Leon Ofman\altaffilmark{3}, Carolus J.~Schrijver\altaffilmark{1},	
        Markus J.~Aschwanden\altaffilmark{1}, Bart De Pontieu\altaffilmark{1}, and Theodore D.~Tarbell\altaffilmark{1}}			



\altaffiltext{1}{Lockheed Martin Solar and Astrophysics Laboratory, 
  Building 252, 3251 Hanover Street, Palo Alto, CA 94304}
\altaffiltext{2}{W.~W.~Hansen Experimental Physics Laboratory, Stanford University, Stanford, CA 94305}
\altaffiltext{3}{Catholic University of America and NASA Goddard Space Flight Center, 
  Code 671, 8800 Greenbelt Road, Greenbelt, MD 20771, USA}

\shorttitle{Fast Mode MHD Waves Observed by {\it SDO} AIA}
\shortauthors{Liu et al.}
\slugcomment{Accepted by ApJ Letters, June 13, 2011}

\begin{abstract}	

Quasi-periodic, propagating fast mode magnetosonic waves
in the corona were difficult to observe in the past
due to relatively low instrument cadences. We report here evidence of 
such waves directly imaged in EUV by the new \sdo AIA instrument.
In the 2010 August 1 C3.2 flare/CME event, we find arc-shaped wave trains 
of 1--5\% intensity variations (lifetime $\sim$200~s) that emanate near the flare kernel
and propagate outward up to $\sim$400~Mm along a funnel of coronal loops.	
Sinusoidal fits to a typical wave train indicate a phase velocity of $2200 \pm 130 \kmps $.	
Similar waves propagating in opposite directions are observed in closed loops 
between two flare ribbons. 
In the $k$--$\omega$ diagram of the Fourier wave power,
we find a bright ridge that represents the dispersion relation 
and can be well fitted with a straight line passing through the origin.
This $k$--$\omega$ ridge shows a broad frequency distribution with
indicative	
power at	
5.5, 14.5, and 25.1~mHz.
The strongest signal at 5.5~mHz (period 181~s)	
temporally coincides with quasi-periodic pulsations of the flare,
suggesting a common origin.		
The instantaneous wave energy flux of $ (0.1$--$2.6)\E{7} \ergs \, \pcms \ps$ 		
estimated at the coronal base is comparable to the steady-state heating requirement of active region loops.

\end{abstract}


\section{Introduction}
\label{sect_intro}



In the last decade, observations from \sohoA, \traceA, \hinodeA, and
ground-based instruments have led to detection of various modes
of magnetohydrodynamic (MHD) waves in the solar corona
(see review by \citealt{Nakariakov.wave-review.2005LRSP....2....3N}),
including (1) oscillations or standing waves of slow modes
\citep{WangTJ.1st-SUMER-slow-mode.2002ApJ...574L.101W, OfmanWang.slow-mode.2002ApJ...580L..85O},
fast kink modes \citep[periods: 2--10~min;][]{Aschwanden.1st-TRACE-wave.1999ApJ...520..880A, 
Schrijver.TRACE-1st-result.1999SoPh..187..261S},
and fast sausage modes \citep[periods: 1--60~s;][]{Nakariakov.sausage-mode-Nobeyama.2003A&A...412L...7N},
and (2) propagating waves of slow modes 
\citep{Ofman.UVCS-polar-hole-wave.1997ApJ...491L.111O, DeForestC.Gurman.slow-wave-plume.1998ApJ...501L.217D,
DeMoortel.TRACE-slow-mode-discover.2000A&A...355L..23D, Ofman.Wang.SOT-wave.2008A&A...482L...9O}
and \Alfven waves (\citealt{Tomczyk.Alfven-wave.2007Sci...317.1192T, 
DePontieu.Alfven-wave-spicule.2007Sci...318.1574D, Cirtain.XRTjet2007Sci...318.1580C,
Okamoto.Alfven-wave.prominence.2007Sci...318.1577O, Jess.Alfven-wave.2009Sci...323.1582J,
LiuW.CaJet1.2009ApJ...707L..37L};	
some of which were interpreted as
kink waves, see \citealt{VanDoorsselaere.kink.or.Alfven.2008ApJ...676L..73V}).


Quasi-periodic propagating fast mode magnetosonic waves 	
with phase speeds $ v_{\rm ph} \sim 1000 \kmps$ in active regions remain the least observed among all coronal MHD waves,
while 	
single-pulse ``EIT waves" \citep{ThompsonB.EIT-wave-discover.1998GeoRL..25.2465T}
of typical speeds $\sim$$200 \kmps$ were interpreted as their quiet Sun counterparts 
(\citealt{Wu.EIT-fast-MHD-wave.2001JGR...10625089W, OfmanThompson.EIT-wave-fast-mode.2002ApJ...574..440O}; 
cf., \citealt{ChenFP.Wu.Coexis-fast-mode.2011ApJ...732L..20C}).
\citet{WilliamsDR.fast-mode-eclipse.2002MNRAS.336..747W} first imaged during an eclipse
a fast wave of $v_{\rm ph}=2100 \kmps$			
in a closed loop.
\citet{Verwichte.fast-mode-Apr21-supra-arcade.2005A&A...430L..65V} later
observed with \trace fast kink modes of $v_{\rm ph}=200$--$700 \kmps$	
in an open-field supra-arcade.	

The scarcity of fast wave observations was mainly due to instrumental limitations.
The new Atmospheric Imaging Assembly \citep[AIA;][]{LemenJ.AIA.instrum.2011SolPhy} 
on the {\it Solar Dynamics Observatory} (\sdoA)
has high cadences up to 10~s, short exposures of 0.1--2~s, and
a $41 \arcmin \times 41 \arcmin$ full-Sun field of view (FOV) at $1\farcs 5$ resolution,
which are all crucial for detecting fast propagating features. 
Within the first year of its launch, AIA has detected 10 quasi-periodic fast propagating (QFP) waves,	
among which the first was mentioned by \citet{LiuW.AIA-1st-EITwave.2010ApJ...723L..53L}
and the best example is presented here.	

\section{Observations and Data Analysis}
\label{sect_obs}


On 2010 August 1, an eruption
\citep{LiuR.2010Aug01.sigmoid.2010ApJ...725L..84L, Schrijver.2010Aug01-long-range-couple.2011JGR}
occurred in NOAA active region 11092, involving 
a coronal mass ejection (CME) and a \goes C3.2 flare that started at 07:25~UT and peaked at 08:57~UT. 	

\subsection{Space-time Analysis}	
\label{subsect_overview}

 \begin{figure*}[thb]      
 \epsscale{0.8}	
 \plotone{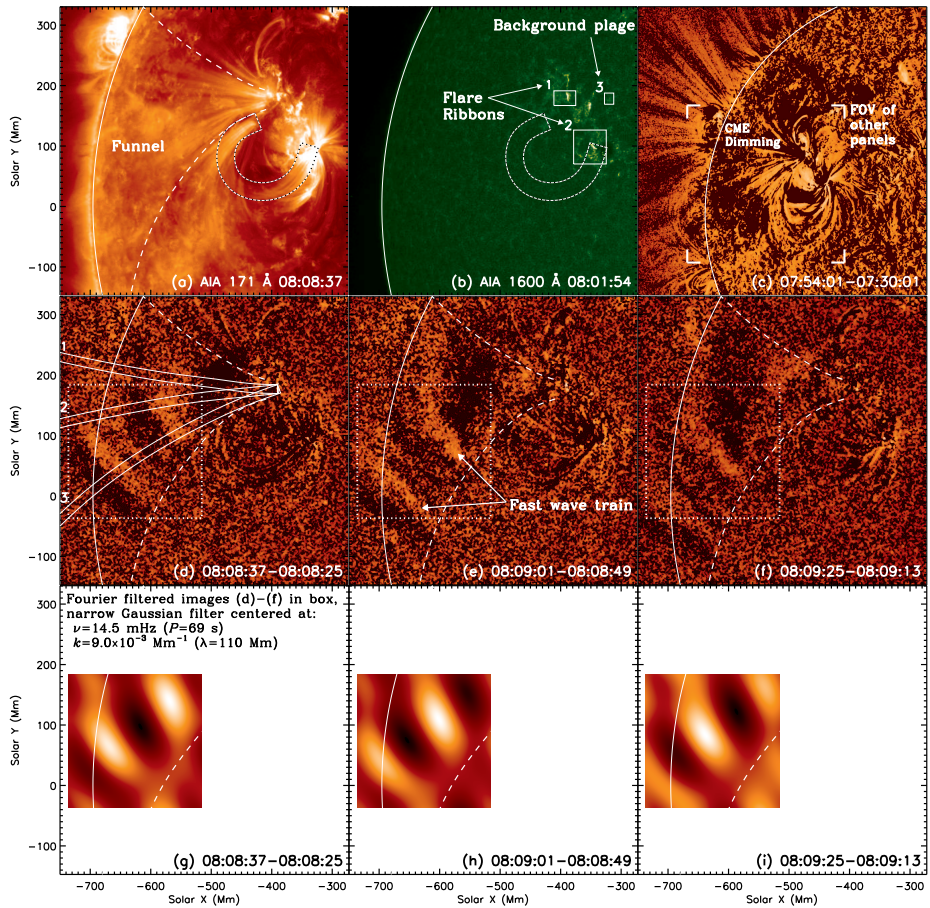}
 \caption[]{\sdo AIA images of QFP waves.	
   (a) 171~\AA\ image (see Animation~1(A)) showing 
 the funnel and loop in which fast waves propagate.
   (b) 1600~\AA\ image (Animation~1(B)) showing flare ribbons.
   (c) 171~\AA\ base difference image (Animation~1(C)) showing dimming behind the CME front.
 The four brackets mark the smaller FOV of the other panels.
   (d)--(f) 171~\AA\ running difference images (Animation~1(D)) showing successive wave fronts
 propagating in the funnel. The three curved cuts	
 are used to obtain space-time diagrams shown in Figure~\ref{timeslice.eps}.	
 The square box marks the region for Fourier analysis in Section~\ref{subsect_fft}.
   (g)--(i) Images of (d)--(f) in the boxed region Fourier filtered with a narrow
 Gaussian centered at the peak in Figure~\ref{power-spec.eps}(b)
 at frequency $\nu=14.5$~mHz ($P=69 \s$) and wave number 
 $k= 9.0\E{-3} \Mm^{-1}$ ($\lambda= 110 \Mm$), which highlight the corresponding
 QFP wave trains (see Animation~1(E) and Section~\ref{subsect_freq-distr}).
 } \label{mosaic.eps}
 \end{figure*}
 \begin{figure*}[thb]      
 \epsscale{1.1}	
 \plotone{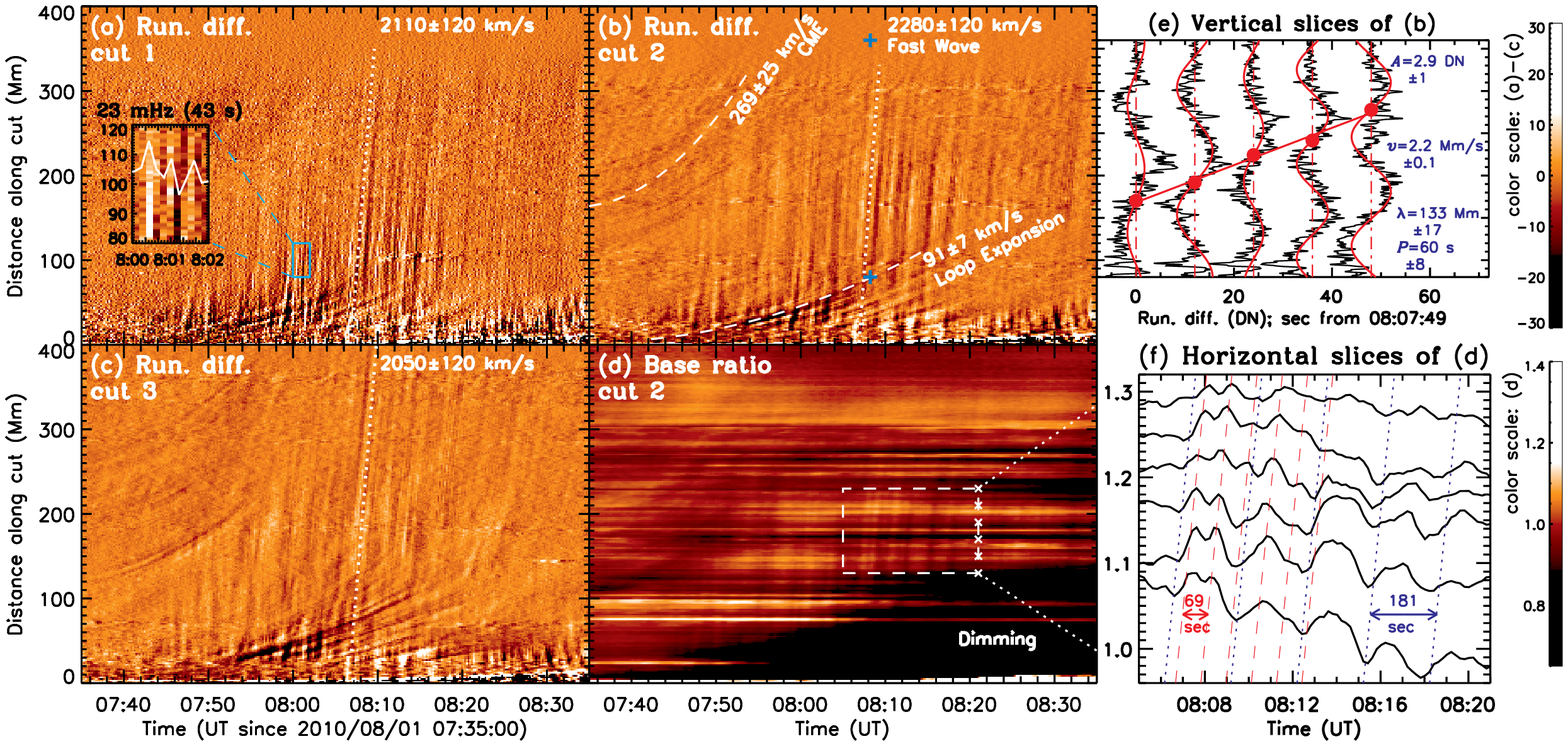}
 \caption[]{Space-time analysis of waves in the funnel.
   (a)--(c) Running difference space-time diagrams obtained from AIA 171~\AA\ images
 along the three cuts shown in Figure~\ref{mosaic.eps}(d).
 The insert in (a) offers an enlarged view for the selected region, overlaid with a distance averaged profile showing a 43~s periodicity.
   (d) Base ratio space-time diagram of cut~2 obtained by normalizing image profiles
 with a pre-event profile.	
 All the space-time diagrams are smoothed with a $3 \times 3$ pixel boxcar,
 except for (a) which is smoothed in space only.
   (e) Vertical slices of (b) at times and distances marked by the two plus signs. 
 They are snapshots of intensity running difference (x-axis) 
 as a function of distance (y-axis) at five consecutive times.	
 Each curve (and thus its average position, 	
 marked by the vertical broken line) 
 is incrementally shifted by $12 \, {\rm DN}$ which equals AIA's $12 \s$ cadence 
 and thus the x-axis also serves as elapsed time. Each profile is fitted with a sine function 
 $A \sin[2\pi(r-r_0) / \lambda]$ shown in red, where $A$ is the amplitude, $\lambda$ wavelength, 
 and $r_0$ the initial phase in distance.  The average fitted parameters and their standard deviations 	
 are listed. The filled circles mark the delayed occurrences at
 the average position, to which a linear fit indicates a phase velocity 
 $v_{\rm ph}=2200 \pm 130 \kmps $.	 
   (f) Horizontal slices of (d) in the selected region, i.e.,
 temporal profiles of intensity base ratio at locations marked by the cross signs.	
 Successive curves at greater distances are shifted upward.		
 The two prominent wave periods of 69 and 181 s
 are marked with slanted lines, indicating wave propagation.	
 } \label{timeslice.eps}
%
 \end{figure*}
%

\subsubsection{Waves in the Funnel}
\label{subsect_wave-in-funnel}

In AIA 171~\AA\ running difference images (Figure~\ref{mosaic.eps}(d)--(f), Animation~1(D))
and even direct and base difference images (Animations~1(A) and 1(C)),
we discovered arc-shaped wave trains emanating near the brightest flare kernel (box~1 in Figure~\ref{mosaic.eps}(b))
and rapidly propagating outward along a funnel of coronal loops that subtend an angle of $\sim$$60 \degree$ near the corona base.
They are successive, alternating intensity variations of 1--5\%, 	
repeatedly launched in the wake of the CME during the rise phase of the flare (07:45--08:45~UT).	
The wave fronts continuously travel beyond the limb, suggesting that they are not
propagating over the solar surface like \citet{Moreton.wave.1960AJ.....65U.494M}
or EIT waves.	
They are	
not observed in the other AIA EUV channels, indicating subtle temperature dependence. 

To analyze wave kinematics, we placed three $20 \arcsec$ (14.7~Mm) wide curved cuts that
start from the brightest flare kernel and follow the shape of the funnel (Figure~\ref{mosaic.eps}(d)).
By averaging pixels across each cut, we obtained image profiles along it
and stacking these profiles over time gives space-time diagrams as shown in 
Figure~\ref{timeslice.eps}, where we see two types of moving features:

(1) The shallow, gradually accelerating stripes represent the expanding coronal loops in the CME
that have final velocities up to $\ge$$260 \kmps$ as indicated by parabolic fits 
(dashed lines in Figure~\ref{timeslice.eps}(b)).
EUV dimming is evident behind these loops (Figures~\ref{mosaic.eps}(c) and \ref{timeslice.eps}(d)), 
indicating evacuation of coronal mass.

(2) The steep, recurrent stripes result from the arc-shaped wave fronts.	
Sinusoidal fits (Figure~\ref{timeslice.eps}(e)) to the spatial profiles along the central cut
yield a projected wavelength $\lambda = 133 \pm 17 \Mm$ and phase velocity $v_{\rm ph}=2200 \pm 130 \kmps $,
giving a period of $P =\lambda / v_{\rm ph}= 60 \pm 8 \s$.
Linear fits to the space-time stripes from the three cuts produced by the same wave front
indicate similar velocities (Figure~\ref{timeslice.eps}(a)--(c)).
(Such velocities measured from projection on the sky plane
are lower limits of their 3D values.)
Each wave front travels up to $\sim$400~Mm with a lifetime of $\sim$200~s before reaching the edge of AIA's FOV,
likely resulting from damping and amplitude decay with distance ($\propto 1/r^2$).

\subsubsection{Waves in Closed Loops}
\label{subsect_wave-in-loop}

 \begin{figure}[thb]      
 \epsscale{0.9}	
 \plotone{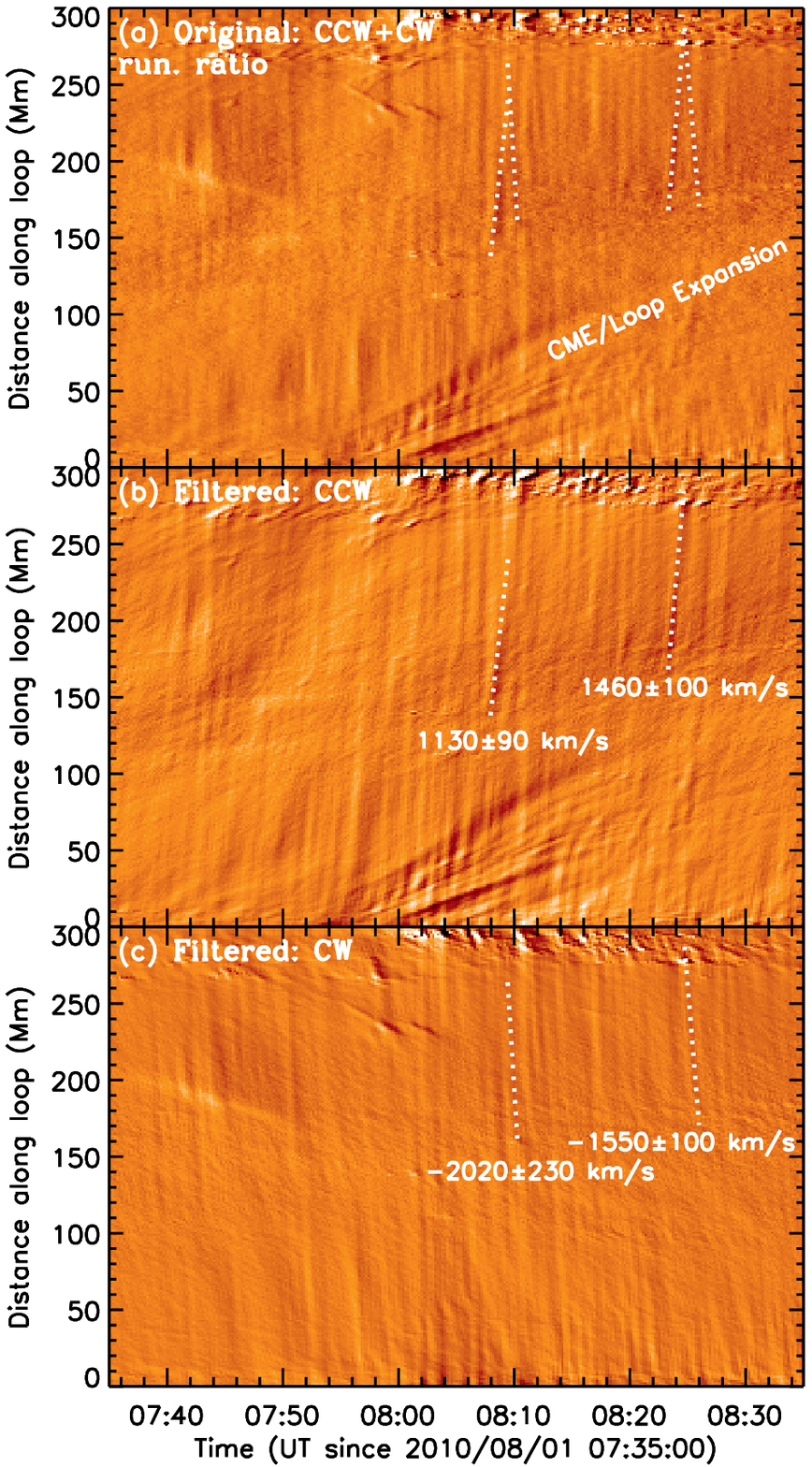}
 \caption[]{
   (a) AIA 171~\AA\ running ratio space-time diagram from the curved cut along coronal loops 
 shown in Figure~\ref{mosaic.eps}(a). Distance is measured from the eastern footpoint.
 Note 	
 fast waves propagating in both counter-clockwise (CCW) and clockwise (CW) directions along the loop. 
   (b) and (c) Fourier filtered version of (a) showing CCW and CW waves separately.
 The linear fits here are repeated in (a).
 } \label{loopslice.eps}
 \end{figure}
At the same time,	
we noticed similar fast propagating waves along closed loops between two flare ribbons
(Figures~\ref{mosaic.eps}(a) and (b)). The space-time diagram (Figure~\ref{loopslice.eps}(a))
from the loop-shaped cut
reveals steep stripes of both positive and negative slopes, particularly near the two footpoints,
which represent waves propagating in opposite directions.
%
The bi-directional propagation can be evidently seen separately in Fourier filtered
space-time diagrams (Figures~\ref{loopslice.eps}(b) and (c); 
see \citealt{TomczykMcIntosh.coronal-time-distance-seism.2009ApJ...697.1384T}).
The linearly fitted phase velocities are similar in the two directions (1000--2000~$\kmps$).
The sudden switches of direction at the western footpoint (top edge of the plot)
near 08:10 and 08:25~UT suggest wave reflection, but a general trend cannot be established.
It is thus not clear whether the bi-directional waves are generated independently, 
or they are the same wave trains reflected repeatedly	
between the footpoints,	


We find no temporal correlation between the waves in the closed loops and those in the funnel
that is dominated by outgoing waves, except for marginal incoming wave signals
near its base (Figure~\ref{timeslice.eps}). Because of their simplicity 
(no superposition of bi-directional propagation), we choose to further
analyze the waves in the funnel with Fourier transform as presented below.

\subsection{Fourier Analysis of Waves in the Funnel}
\label{subsect_fft}

%

\subsubsection{Overall $k$--$\omega$ Diagram}
\label{subsect_overall-k-omega}

 \begin{figure*}[thbp]      
 \epsscale{1.05}	
 \plotone{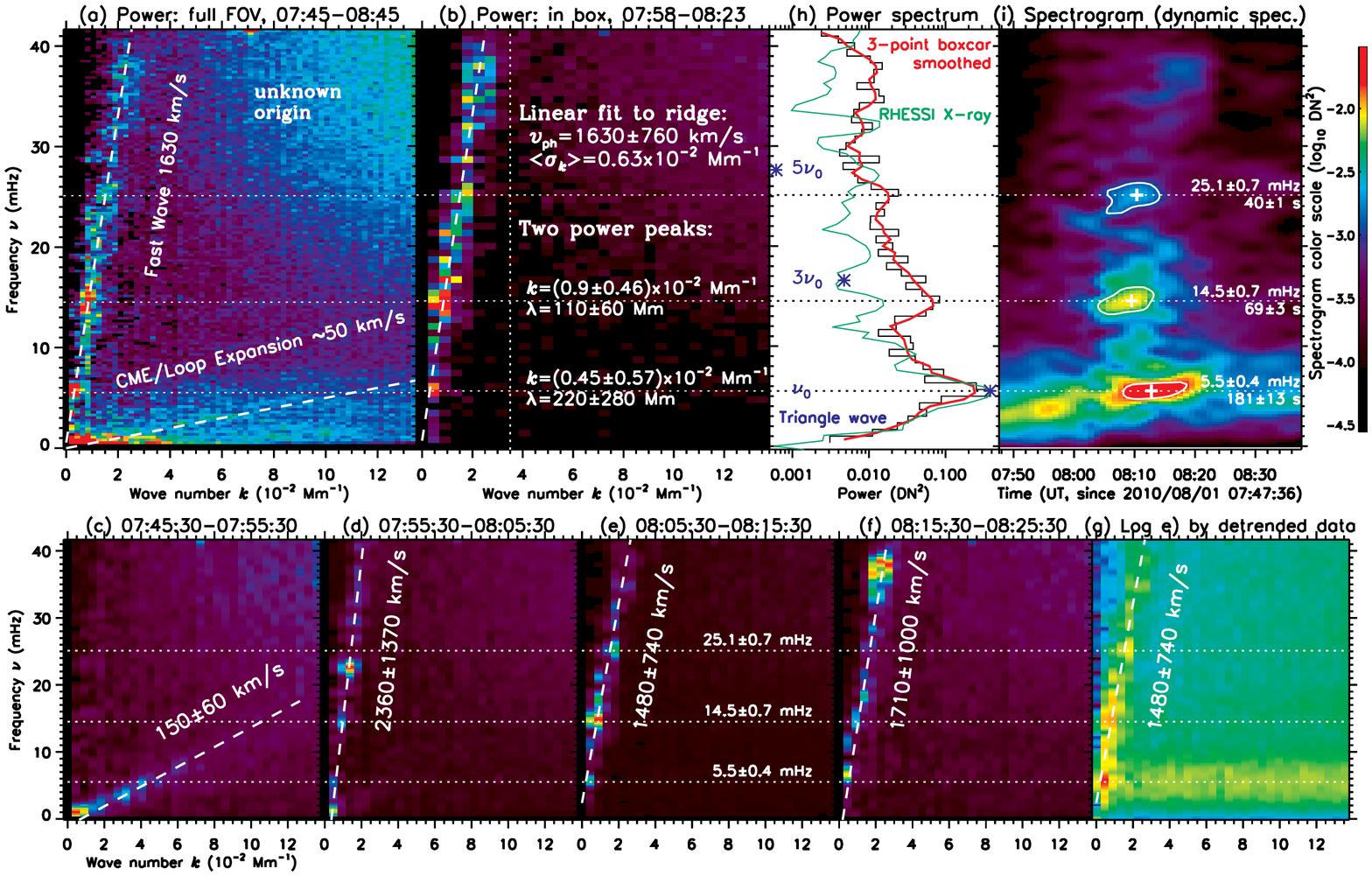}
 \caption[]{Fourier analysis of wave properties.
  (a) Fourier power ($k$--$\omega$ diagram) of a 3D data cube of 171~\AA\ running difference images
 during 07:45--08:45~UT in the full FOV of Figure~\ref{mosaic.eps}(a).
  (b) Same as (a) but for 07:58--08:23~UT in the boxed region of Figure~\ref{mosaic.eps}(d).
 The white dashed line (repeated in (a)) is a power-weighted linear fit		
 to the $(k,\nu)$ positions of pixels greater than 
 10\% of the maximum power in the $k \le k_{\rm max}$ range 
 (marked by the vertical dotted line). 		
  (c)--(f) Same as (b) but for images masked with a running time window
 whose FWHM is labeled on each panel	
 (see Animation~4).
  (g) Same as (e) but on log scale from detrended (rather than running difference) images (see Section~\ref{subsect_freq-distr}).
 The diffuse horizontal band is an artifact at the 5~mHz detrending cutoff frequency
 and is $\sim$5\% of the QFP wave power here.
  (h) Power spectrum vs.~frequency obtained by averaging in 	
 $k \le k_{\rm max}$ on a $k$--$\omega$ diagram that is the same as (b) but from detrended images.
  (i) Spectrogram obtained by compiling wave number averaged power
 as shown in (h) from $k$--$\omega$ diagrams at different times as shown in (g).	
 The x-axis here refers to central times of the running window.
 Prominent ``islands"  are contoured at the 50\% level; their peaks are 
 marked by plus signs and the peak frequencies (periods) by horizontal dotted lines.
 The frequency uncertainties are the standard deviations within the contours.
 } \label{power-spec.eps}	
 \end{figure*}
%
We extracted a 3D data cube in $(x, y, t)$ coordinates, i.e., a time series of 171~\AA\ running difference images
for the FOV of Figure~\ref{mosaic.eps}(a) during 07:45--08:45~UT.
We obtained the Fourier power of the data cube
on the $(k_x, k_y, \nu)$ basis of wave number $k_x$
and $k_y$ and frequency $\nu$. We then summed the power 	
in the azimuthal $\theta$ direction of cylindrical coordinates $(k, \theta, \nu)$, where $k= \sqrt{k_x^2 + k_y^2}$
\citep[e.g.,][]{DeForestC.TRACE-1600-fast-mode.2004ApJ...617L..89D}.
This yields a $k$--$\omega$ diagram of wave power 	
at a resolution of $\Delta k= 2.09 \E{-3} \Mm^{-1}$ and $\Delta \nu= 0.277 {\rm mHz}$
as shown in Figure~\ref{power-spec.eps}(a).
%
We find a steep, narrow ridge that describes the dispersion relation of 
the fast propagating waves, together with a shallow, diffuse ridge that represents
those slowly expanding loops at velocities on the order of $50 \kmps$. 

To isolate the fast propagating waves (at the expense of reduced frequency resolution),
we repeated this analysis for a smaller boxed region
as shown in Figure~\ref{mosaic.eps}(d) and a shorter duration of 07:58--08:23~UT
in which these waves are prominent. The resulting $k$--$\omega$ diagram (Figure~\ref{power-spec.eps}b)
better shows the steep ridge that can be fitted with a straight line passing through the origin.
This gives average phase ($v_{\rm ph}= \nu/ k$) and group ($v_{\rm gr}= d\nu / dk$) velocities
of $1630 \pm 760 \kmps $, 	
which cannot be distinguished
in the observed range up to the Nyquist frequency of $41.7$~mHz given by AIA's 12~s cadence due to the large uncertainty. 

\subsubsection{Temporal Evolution of $k$--$\omega$ Diagram}
\label{subsect_k-omega-evolve}

We repeated the above procedure for a data cube of the boxed region during 07:45--08:45~UT 
masked with a running time window that has a full width half maximum (FWHM) of 10 minutes with cosine bell tapering on both sides. 
We shifted the window by 1~min at a time 
(only 6 such windows are independent	
in the 1~hr duration) and obtained a corresponding $k$--$\omega$ diagram,
as shown in Figures~\ref{power-spec.eps}(c)--(f) and Animation~4.
The early $k$--$\omega$ diagrams are dominated by a shallow ridge with an increasing slope 		
that indicates the CME acceleration. When the CME front moves out of the FOV, 
a steep ridge corresponding to the fast propagating waves becomes progressively evident
with a slope varying in the 1000-2000~$\kmps$ range.
 
\subsubsection{Frequency Distribution of Fourier Power}
\label{subsect_freq-distr}

We note that running difference (time derivative) in images used above, similar to a highpass filter,
essentially scales the original signal with frequency $\nu$ and applies a $\nu^2$ factor to the Fourier power.
To recover the intrinsic power amplitude, we replaced running difference images
with detrended images obtained by subtracting images running-smoothed in time with a 
200~s boxcar, introducing a low-frequency cutoff of 5~mHz that is below all strong peaks on the ridge in Figure~\ref{power-spec.eps}(b).		
We then repeated the above analysis in Sections~\ref{subsect_overall-k-omega} and \ref{subsect_k-omega-evolve}.
The new $k$--$\omega$ diagrams (e.g., Figure~\ref{power-spec.eps}(g) vs.~(e))
exhibit the expected general trend of decreasing power with frequency,
and as a result the steep ridge becomes less evident at high frequencies.

We averaged the new version (not shown) of the overall $k$--$\omega$ diagram of Figure~\ref{power-spec.eps}(b) in 
wave number and obtained a power spectrum for the QFP waves (Figure~\ref{power-spec.eps}(h)).	
We repeated this for the new $k$--$\omega$ diagrams at different times (e.g., Figure~\ref{power-spec.eps}(g))
and compiled 	
a running spectrogram 	
(Figure~\ref{power-spec.eps}(i)). 	
The waves display a broad frequency distribution \citep[cf.,][]{Tomczyk.Alfven-wave.2007Sci...317.1192T},
with power peaks of ratio 1: 1/4.6 : 1/15.8  	
($\propto \nu ^{-1.8 \pm 0.2}$)	
at frequencies $\nu = 5.5 \pm 0.4$, $14.5 \pm 0.7$, and $25.1 \pm 0.7$~mHz		
of ratio $1: (2.6 \pm 0.2) : (4.5 \pm 0.3)$.		
For comparison, a triangle wave of $\nu_0=$5.5~mHz has non-zero Fourier power
(blue asterisks, Figure~\ref{power-spec.eps}(h)) at frequencies of similar ratio $1:3:5$ 	
that drops faster with $\nu^{-4}$.



The Fourier power from running difference and detrended images yields
consistent peak frequencies,	
which can be visually identified in the space-time domain. 
The lowest frequency $\nu_0=5.5$~mHz ($P_0=181 \pm 13 \s \approx 3~{\rm min}$) 
manifests as slow modulations 	
in Figures~\ref{timeslice.eps}(b) and (d) at 08:06--08:18~UT.	
The next period $69 \pm 3 \s$ (14.5~mHz), dominating the power from running difference images (Figure~\ref{power-spec.eps}(b)),
matches the temporal spacing between bright stripes near 08:08~UT in Figures~\ref{timeslice.eps}(a)--(c)
and the period given by the sinusoidal fits (Figure~\ref{timeslice.eps}(e)).	
The corresponding wave fronts are prominent in the original and Fourier filtered images
(Figures~\ref{mosaic.eps}(d)--(i), Animation~1(E)).	
These two periods are also evident in the emission profiles of Figure~\ref{timeslice.eps}(f).
The higher frequency $25.1$~mHz ($40 \pm 1 \s$) has considerably weaker power and
a close frequency of 23~mHz (see Figure~\ref{power-spec.eps}(d)) can be seen in the spacing 
of narrow stripes near 08:01~UT (Figure~\ref{timeslice.eps}(a)),		
when the other two frequencies are not yet strong.		


\subsection{Common 3~min Periodicity in Waves and Flare}		
\label{subsect_flare}
 \begin{figure}[thbp]      
 \epsscale{1.1}	
 \plotone{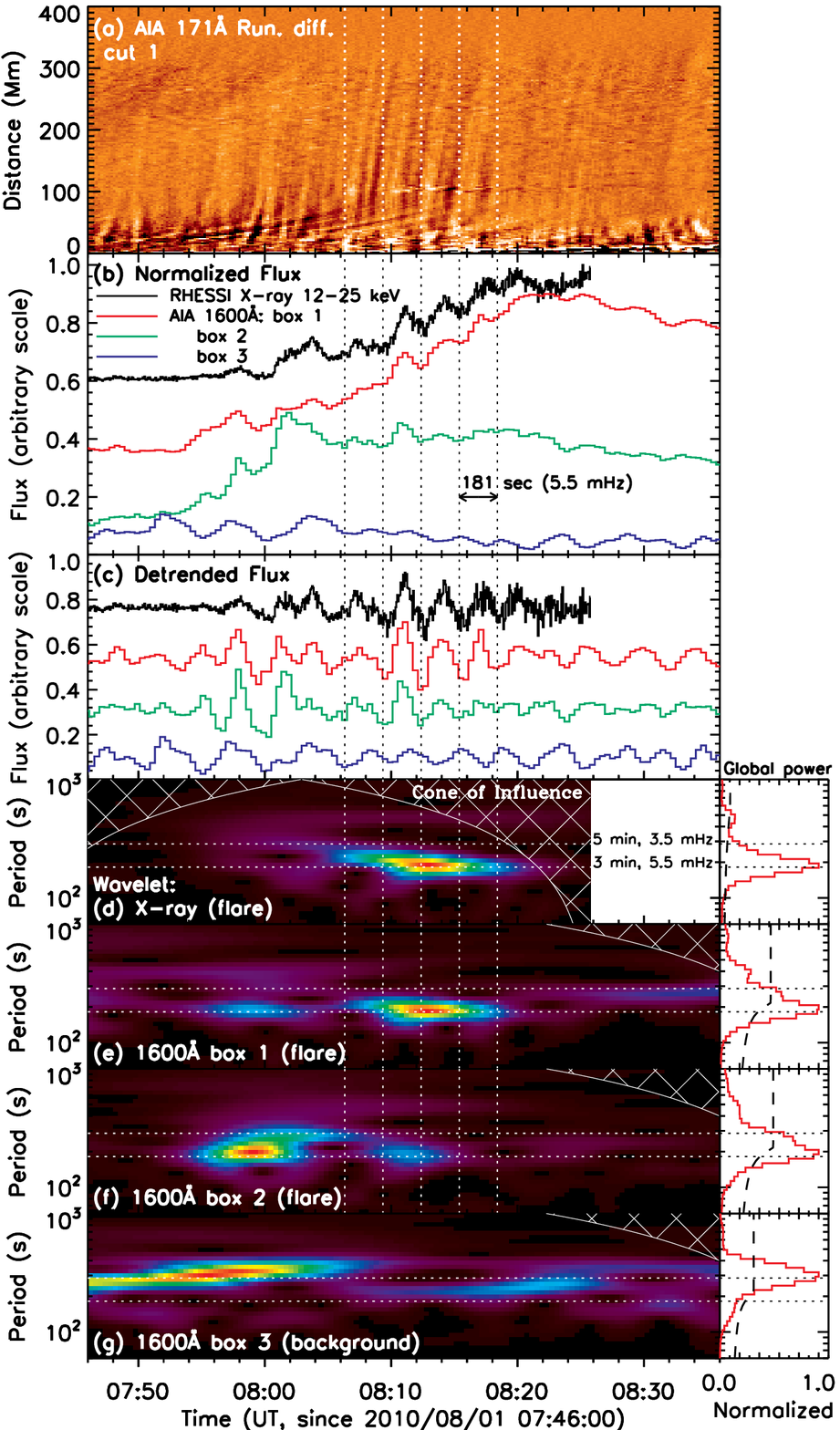}
 \caption[]{Quasi-periodic flare pulsations.
   (a) Same as Figure~\ref{timeslice.eps}(a) but smoothed with a $3 \times 3$ boxcar.
   (b) \hsi 12-25 keV X-ray flux and 
 AIA 1600~\AA\ fluxes integrated over the boxes shown in Figure~\ref{mosaic.eps}(b).	
   (c) Same as (b) but for detrended fluxes obtained by subtracting smoothed fluxes using a 200~s boxcar.
   (d)--(g) Wavelet and global power 		
 of the curves in (c), but for extended durations to reduce the hatched cone of influence 	
 of edge effects \citep{Torrence.Compo.Wavelet.1998BAMS...79...61T}.
 The dashed line in the global power indicates the 95\% significance level.
 The horizontal dotted lines mark the 3.5 and 5.5~mHz frequencies (periods: $\sim$5 and 3~min).
 The vertical dotted lines mark the onsets of the 181~s modulations
 shown in Figure~\ref{timeslice.eps}(f).
 %
 } \label{flare-oscil.eps}
 \end{figure}
%
As shown in Figure~\ref{flare-oscil.eps}(b),
the \hsi X-ray flux 	
and AIA 1600~\AA\ fluxes of flare ribbons 
(particularly the brightest one in box~1 where the funnel is rooted, see Figure~\ref{mosaic.eps}(b) and Animation~1(B)) 
exhibit bursty bumps at a 3~min period (5.5~mHz). 
The onsets of these pulsations (vertical dotted lines)		
coincide with those of the slow modulations on the QFP waves	
 (Figure~\ref{flare-oscil.eps}(a)).
This can be also seen in the wavelet power of these flare emissions (Figures~\ref{flare-oscil.eps}(d)--(g)).
The Fourier power of the X-ray flux (green curve, Figure~\ref{power-spec.eps}(h))
is consistent with that of the QFP waves $\lesssim$10~mHz, but significantly lower at higher frequencies.
It also matches that of the triangle wave up to the third harmonics because of its triangular pulse shape (Figure~\ref{flare-oscil.eps}(c)).
In contrast, the 1600~\AA\ flux of a background plage (box~3 in Figure~\ref{mosaic.eps}(b)) 
is constantly dominated by the 5~min (3.5~mHz) photospheric p-mode oscillations
(Figure~\ref{flare-oscil.eps}(b) and (f)).

\subsection{Estimate of Wave Energy and Magnetic Field}
\label{subsect_energy}

The energy flux carried by the QFP waves can be estimated with the kinetic energy
of the perturbed plasma,		
 $ E = \rho (\delta v)^2 v_{\rm ph} /2
     \geq \rho ( \delta I / I)^2 v_{\rm ph}^3 / 8  $ 
\citep{AschwandenM.wave.energy.2004ESASP.575...97A},
where we have assumed that the observed intensity variation $\delta I$ results from
density modulation $\delta \rho$ and 
used $\delta v/v_{\rm ph} \geq \delta \rho / \rho = \delta I / (2I)$ for magnetosonic waves since $I \propto \rho^2$.  
If we take $v_{\rm ph} = 1600 \kmps $ and $\delta I / I=$ 1\%--5\%
observed in the mid-range of the funnel (200~Mm from the flare kernel),
and use the corresponding number density $n_e \gtrsim 10^8 \pcmc$ 
estimated with the 171~\AA\ channel response (following \citealt{DePontieu.hot-corona-origin.2011Sci...331...55D}), 
we reach an energy flux $E \gtrsim (0.1$--$2.6)\E{5} \ergs \pcms \ps$.
The diameter of the funnel here has increased $\sim$10 times
from the coronal base, where the the wave energy flux shall be $\gtrsim$$10^2$ times higher
by continuity of energy flow, if we assume the waves being generated there and consider damping on their path. 	
This energy flux is more than sufficient for heating the {\it local} active region loops
\citep{Withbroe.Noyes.coronal-heating-flux.1977ARA&A..15..363W}.
However, considering the limited temporal and spatial extent of these waves,
they are unlikely to play an important role in heating the quiescent {\it global} corona.

Assuming the measured phase speed $v_{\rm ph}$ equal to the fast mode speed along
magnetic field lines in the funnel, which is the \Alfven speed $v_A = B / \sqrt{4 \pi \rho}$, the magnetic field
strength is estimated as $B=v_{\rm ph} \sqrt{4 \pi \rho} \gtrsim 8 \gauss$.

\section{Discussion}
\label{sect_summ_100801}

We propose that these QFP waves imaged with AIA's unprecedented capabilities
are fast mode magnetosonic waves that have been theoretically predicted and simulated 
(e.g., \citealt{Bogdan.coronal-wave-simul.2003ApJ...599..626B, Fedun.magnetosonic-simul.2011ApJ...727...17F}; Ofman et al.~in preparation),
but rarely observationally detected. We speculate their possible origin as follows.

\begin{enumerate}

\item
The accompanying CME is unlikely to be the wave trigger, because it takes place gradually for $\sim$30~min 
($\gg$ wave periods, Figure~\ref{timeslice.eps}(b)) and its single pulse would be difficult to 
sustain oscillations lasting $\sim$1~hr as observed here without being damped.
However, the environment in its wake might be favorable for these waves.

\item
The common 3~min periodicity (Section~\ref{subsect_flare}) of the QFP waves and flare quasi-periodic pulsations 
\citep[QPPs;][]{Nakariakov.Melnikov.QPP.2009SSRv..149..119N, Kupriyanova.QPP.2010SoPh..267..329K}
suggests a common origin. Quasi-periodic magnetic reconnection and energy release 
can excite both flare pulsations \citep{OfmanSui.HXR-oscil.2006ApJ...644L.149O, Fleishman.QPP-periodic-acc.2008ApJ...684.1433F} 
and MHD oscillations that drive QFP waves, or in turn, MHD oscillations responsible for the waves can modulate 
energy release and flare emission \citep{Foullon.QPP.kink-mode.2005A&A...440L..59F}.
This periodicity is the same as that of 3~min chromospheric oscillations, further suggesting their possible modulation 
on reconnection \citep{Chen.Priest.P-mode-driven-reconn.2006SoPh..238..313C,
Heggland.wave-driven-reconn.2009ApJ...702....1H,	
McLaughlin.wave-modulate-X-reconn.2009A&A...493..227M}.

However, the deficit of flare power at higher wave frequencies ($\gtrsim$10~mHz, Figure~\ref{power-spec.eps}(h)) is somewhat puzzling.
Perhaps the waves are driven by a multi-periodic exciter that produces no detectable flare signals 
at these frequencies. A future study of similar events will further shed light on the nature
of these waves.

\end{enumerate}




\acknowledgments
{This work was supported by AIA contract NNG04EA00C. 
LO was supported by NASA grants NNX08AV88G and NNX09AG10G.
Wavelet software	
was provided by C.~Torrence and G.~Compo.
We thank Nariaki Nitta for helpful discussions.
}



{\scriptsize

}



\end{document}